\documentclass[journal]{IEEEtran}
\ifCLASSINFOpdf
\else
\fi
\usepackage{amsmath}
\usepackage{amssymb}
\usepackage{graphicx}
\usepackage{epstopdf}
\usepackage{color}
\usepackage{hyperref}  
\usepackage{booktabs}  
\usepackage{multirow}  
\usepackage{cite}      
\usepackage{verbatim}  
 \allowdisplaybreaks[4] 
\usepackage{setspace}  

\renewcommand{\t}{{\sf \tiny T}}

\begin{document}
\newtheorem{theorem}{Theorem}
\newtheorem{remark}{Remark}
\newtheorem{corollary}{Corollary}
\newtheorem{definition}{Definition}
\newtheorem{problem}{Problem}
\newtheorem{lemma}{Lemma}
\newtheorem{assumption}{Assumption}

\title{ Full-State Prescribed Performance-Based Consensus of Double-Integrator Multi-Agent Systems with Jointly Connected Topologies }
%
%
%

\author{Yahui Hou and Bin Cheng, Member, IEEE
\thanks{  Yahui Hou is with the School of Automation, Central South University, Changsha 410083, China
         (e-mail: yahuihou@163.com).

         Bin Cheng is with the Department of Control Science and Engineering,
College of Electronics and Information Engineering, Tongji University,
Shanghai 201804, China (e-mail: bincheng@tongji.edu.cn).  }
\thanks{Corresponding authors: Yahui Hou and Bin Cheng. }
        }

%
%

\markboth{Jan~2024 }%
{Shell \MakeLowercase{\textit{et al.}}: Bare Demo of IEEEtran.cls for IEEE Journals}
%



\maketitle

\begin{abstract}
This paper addresses the full-state prescribed performance-based consensus problem for double-integrator multi-agent systems with jointly connected topologies.
To improve the transient performance, a distributed prescribed performance control protocol consisting of the transformed relative position and  the transformed relative velocity is proposed, where the communication topology satisfies the jointly connected assumption.
Different from the existing literatures, two independent transient performance specifications imposed on relative positions and relative velocities can be guaranteed simultaneously.
A numerical example is ultimately used to validate the effectiveness of proposed protocol.
\end{abstract}


%
\IEEEpeerreviewmaketitle

\section{Introduction}
%
%
%
%


Recently, multi-agent systems (MASs) have received considerable attention in completing a range of tasks such as rescue, transportation, and exploration, mainly because they are efficient and inexpensive\cite{2020Networked}.
More and more research topics were conducted on the cooperative control of MASs, in which consensus reaching is a classical topic\cite{2017Recent}.
The consensus of MASs was achieved for different dynamics, including single-integrator dynamic \cite{2004Consensus}, double-integrator dynamic \cite{Yu2013Consensus, zeng2022second}, linear dynamic \cite{ren2021optimal, sun2023consensus}, and nonlinear dynamic \cite{tan2020distributed}.

Although the control protocols in literatures \cite{2004Consensus, Yu2013Consensus, zeng2022second, ren2021optimal, sun2023consensus, tan2020distributed} achieved asymptotic convergence as expected, the steady and transient-state performance during the convergence process were usually neglected.
To address this issue, the prescribed performance control (PPC) method based on performance functions
and error transformation functions was first proposed in \cite{2008Robust}, where the tracking errors were confined within an arbitrarily small residual set, with predefined minimum convergence rate and maximum overshoot.
Soon afterwards, the PPC method was applied to the single-integrator and double-integrator MASs to achieve consensus \cite{Karayiannidis2012Multi, Macellari2017Multi},
in which the trajectories of relative positions defined as the difference among the connected agents respected the performance bound.

In detail, we then review the application of PPC method in MASs.
In \cite{Macellari2017Multi}, Macellari et al. proposed a distributed control protocol consisting of a linear combination term between relative positions and relative velocities as well as an additional absolute velocity term, such that the performance specifications imposed on the above linear combination term can be guaranteed.
The performance specifications are imposed on the linear combination term, however, relative positions and relative velocities cannot be confined within their respective performance bounds.
In terms of the controller designed in \cite{Macellari2017Multi}, modulating the performance specifications on relative velocities makes the stability analysis more complicated.
In \cite{Karayiannidis2016AModel}, the designed controller guaranteed a transient performance for the robot joint position and velocity tracking errors.
Recently, more research results on this topic can be found in \cite{chen2019consensus, chen2019Second, Hou2021Event, Hou2023Event}.

Up to this point, no research results were found in which the relative positions and relative velocities among connected agents evolve within their respective predefined performance bounds.
In short, these independent performance specifications associated with positions and velocities form the so-called full-state prescribed performance for double-integrator MASs.
The above literatures mainly considered fixed topology, while switching topology cases caused by the failures and changes of communication link among neighboring agents are rarely involved \cite{cheng2018event, wan2023differentially}.

Motivated by the above discussion, the full-state prescribed performance-based consensus control problem for double-integrator MASs with jointly connected topologies is addressed.
To be specific, a vital advantage of this paper, is that a novel distributed control protocol is designed to improve the transient performance of both relative positions and relative velocities, which are different from the existing PPC methods based on MASs with single-integrator \cite{Karayiannidis2012Multi, chen2019consensus} and double-integrator \cite{Macellari2017Multi, Hou2021Event, Hou2023Event} dynamics.
Furthermore, considering the jointly connected topology case, all agents are driven to converge asymptotically by applying prescribed performance-based control protocol, and the design parameters are independent of topology information.
It is worth mentioning that topology graphs are no longer limited to tree and connected graph cases \cite{chen2019consensus, chen2019Second, Hou2021Event}, and MASs can achieve the accurate consensus rather than practical consensus \cite{Karayiannidis2016AModel, chen2019Second}.


\section{Background and Problem Formulation}

\subsection{ Notations and Graph Theory}

For convenience, we summarize some notations as follows.
  $\mathbb{R}^N$: the set of $N$ dimension real column vectors;
  $I_N$: the $N \times N $ identity matrix;
    $x^{\t}$: the transposition of $x$;
  $ {\rm diag}\{d_1,d_2,\cdots,d_n\} $: diagonal matrix with entries $ d_1,d_2,\cdots,d_n $.

An undirected graph $ \mathcal{G}_{\sigma(t)} $ is composed of the node set $ \mathcal{V}=\{1,2,\cdots,N\} $ and edge set $ \mathcal{E}_{\sigma(t)} \subset \mathcal{V} \times \mathcal{V} $,
where $ \sigma(t): [0, +\infty ) \to \mathcal{P} $, $ \mathcal{P} = \{ 1,2, \cdots, n_0 \} $  is a piecewise constant switching signal.
Let $ \mathcal{A}_{\sigma(t)} = [a_{ij}(t)] \in \mathbb{R}^{N \times N} $ be the adjacency matrix where $ a_{ij}(t) = 1 $ if $ (i,j) \in \mathcal{E}_{\sigma(t)} $ and $ a_{ij}(t) = 0 $ otherwise.
The neighbor set of node $i$ is $\mathcal{N}_i(t) = \{ j \in \mathcal{V}: (i,j) \in \mathcal{E}_{\sigma(t)} \}$.
The Laplacian matrix is defined as $ \mathcal{L}_{\sigma(t)}= \mathcal{D}_{\sigma(t)} - \mathcal{A}_{\sigma(t)} $,
where degree matrix $ \mathcal{D}_{\sigma(t)} = {\rm diag} \{ d_1(t), d_2(t), \cdots, d_N(t) \}$ with diagonal entry $ d_i(t) = \sum_{j\in \mathcal{N}_i(t)} a_{ij}(t) $.
The incidence matrix is defined as $B_{\sigma(t)}=[b_{il}(t)]\in \mathbb{R}^{N\times  M_{\sigma(t)} }$, $i \in \mathcal{V}$, $l \in \mathcal{H}_{\sigma(t)}$ where $\mathcal{H}_{\sigma(t)} = \{1, \cdots, M_{\sigma(t)} \}$ is a edge set.
Besides, edge $(i,j) \in \mathcal{E}_{\sigma(t)} $, $ j\in \mathcal{N}_i(t) $ is described as edge $l \in \mathcal{H}_{\sigma(t)} $.
For the connected graph, the edge Laplacian matrix $\mathcal{L}_{\sigma(t)}^e = B_{\sigma(t)}^{\t}B_{\sigma(t)} $ is positive semidefinite \cite{Hou2021Event, Hou2023Event}.

Suppose there exists an infinite, bounded and contiguous time interval $[t_k,t_{k+1})$, $k \in \mathbb{N}$ 
where $t_0 = 0$ and $ t_{k+1} - t_k \leq \nu $ for positive constant $\nu$.
During each interval $[t_k,t_{k+1})$, there is a sequence of nonoverlapping subintervals
 $ [t_k^0,t_{k}^1), \cdots, [t_k^q,t_{k}^{q+1}), \cdots, [t_k^{m_k-1},t_{k}^{m_k}) $,
 with $ t_k^0 = t_k $ and $ t_{k}^{m_k} = t_{k+1} $, satisfying $t_{k}^{q+1} - t_k^q \geq \tau > 0$,
 $ 0 \leq q \leq m_k $, and $\tau$ is called dwell time.
 The graph $\mathcal{G}_{\sigma(t)}$ is fixed for $t \in [t_k^q,t_{k}^{q+1})$.
 If the union graph $ \cup_{q=0}^{m_k} \mathcal{G}_{\sigma(t_k^q)} $ is connected,
   then $\mathcal{G}_{\sigma(t)}$ is jointly connected.

\subsection{Prescribed Performance Control} \label{Prescribed Performance Control}

Partly referring to \cite{2008Robust, Karayiannidis2012Multi}, the prescribed performance is achieved if element $s_l(t)$ of tracking errors $s(t) \in \mathbb{R}^{M_{\sigma(t)}} $ evolves within the following performance bounds
\begin{align} \label{prescribed performance}
-\rho_{s,l}(t)  < s_l(t) < \rho_{s,l}(t), ~ \forall l \in \mathcal{H}_{\sigma(t)}
\end{align}
where performance function $\rho_{s,l}(t) = ( \rho_{s,l}^0 - \rho_{s,l}^\infty ) e^{- \mathfrak{g} t } + \rho_{s,l}^\infty$ with positive constants $\rho_{s,l}^0$, $\rho_{s,l}^\infty$ and $\mathfrak{g}$.
Next, we normalize $s_{l}(t)$ by
$ \hat{s}_{l}(t) = \frac{s_{l}(t)}{{\rho_{s,l}(t)}}$,
and define the prescribed performance regions
$D_{\hat{s}_{l}} \triangleq \left\{ \hat{s}_{l}(t)\in {\mathbb R}: \hat{s}_{l}(t) \in (-1, 1) \right\}$
with $ D_{\hat{s}} \triangleq D_{\hat{s}_1} \times D_{\hat{s}_2} \times \cdots \times D_{\hat{s}_{M_{\sigma(t)}}} $.
Let $ \hat{s}= [\hat{s}_1, \cdots, \hat{s}_{M_{\sigma(t)}}]^{\t}$ and
$\rho_{s}(t)={\rm diag} \{ \rho_{s,1}(t), \cdots,\rho_{s,{M_{\sigma(t)}}}(t) \}$.
Thus $ \hat{s}_l (t) \in D_{\hat{s}_l} $, $ t \geq 0 $ is equivalent to \eqref{prescribed performance}.

The normalized value $\hat{s}_{l}(t)$ is transformed by tracking error transformation functions that define a smooth and strictly increasing mapping $\varepsilon_{s,l}: D_{\hat{s}_{l}} \to (-\infty , \infty )$, with $\varepsilon_{s,l}(0)=0$.
More specifically, one can select transformation function as
\begin{align} \label{transformation func-1}
\varepsilon_{s,l}(\hat{s}_{l})= \ln \bigg( \frac{ 1 +\hat{s}_{l}}{ 1-\hat{s}_{l} } \bigg).
\end{align}
The time derivative of $\varepsilon_{s,l}(\hat{s}_{l})$ is
\begin{align} \label{dot_vare}
\dot{\varepsilon}_{s,l} (\hat{s}_{l})
 = J_{s,l}(\hat{s}_{l},t)\left[ {\dot{s}_{l}  + \alpha_{s,l}(t)s_{l} } \right]
\end{align}
%
with $ J_{s,l}(\hat{s}_{l},t) = \frac{ 2 }{ 1-\hat{s}_{l}^2 }  > 0 $ and $\alpha_{s,l}(t) = - \frac{ \dot{\rho}_{s,l}(t) }{ \rho_{s,l}(t)}$.
Let $ \alpha_{s}(t) = {\rm diag} \{ \alpha_{s,1}(t), \cdots, \alpha_{s,{M_{\sigma(t)}}}(t)\} $.
By calculating, the following inequality
\begin{align}\label{baralpha}
\sup_{t\geq 0, l \in \mathcal{H}_{\sigma(t)}}[\alpha_{s,l}(t) ] \leq \bar{\alpha}_s
\end{align}
holds for a positive constant $\bar{\alpha}_s$.
In view of the properties of transformation function $ \varepsilon_{s,l} (\hat{s}_{l}) $,
with constants $\zeta_{s,1}, \zeta_{s,2} > 0$, the following inequality
\begin{align}\label{ineq vare}
 s  J_{s }(\hat{s},t) \varepsilon_{s}(\hat{s}) \ge \zeta_{s,1} \varepsilon_{s}^2(\hat{s}), ~
 s  J_{s }(\hat{s},t) \varepsilon_{s}(\hat{s}) \ge \zeta_{s,2}  \hat{s}^2
\end{align}
holds\cite{Karayiannidis2016AModel},
where
$  s =  [ s_1, \cdots, s_{M_{\sigma(t)}}]^{\t} $,
$  \hat{s} =  [ \hat{s}_1, \cdots, \hat{s}_{M_{\sigma(t)}}]^{\t} $,
$ \varepsilon_{s} ( \hat{s} ) =  [ \varepsilon_{s,1}(\hat{s}_{1}), \cdots, \varepsilon_{s,{M_{\sigma(t)}}}(\hat{s}_{M_{\sigma(t)}}) ]^{\t} $,
and
$ J_{s}( \hat{s},t ) = {\rm diag} \{ J_{s,1}(\hat{s}_1,t) , \cdots, J_{s,{M_{\sigma(t)}}}(\hat{s}_{M_{\sigma(t)}},t)  \} $.

\begin{lemma}[\cite{stamouli2022robust}] \label{lemma-1}
Given any positive constants $k$ and $D$, we have
  \begin{align*}
    -k \varepsilon_s^{\t}(\hat{s})J_s(\hat{s},t)s  + D \leq 0
  \end{align*}
for all $ \|\varepsilon_s(\hat{s})\| > \bar{\varepsilon}_s $, with constant $\bar{\varepsilon}_s > 0$.
\end{lemma}

\subsection{Problem Formulation}

Consider a MAS with $N$ agents modeled by double-integrator dynamic
\begin{equation}\label{mas}
\dot{x}_i(t) = v_i(t), \  \dot{v}_i(t) = u_i(t), \  i\in\mathcal{V}
\end{equation}
where ${x}_i(t), v_i(t), u_i(t) \in \mathbb{R} $ represent the position, velocity and control input of agent $i$, respectively.

\begin{assumption} \label{assumption-2}
The graph $\mathcal{G}_{\sigma(t)}$ are jointly connected across each interval $[t_k, t_{k+1})$, $k \in \mathbb{N}$.
\end{assumption}

Based on the matrix $B_{\sigma(t)}$,
we define relative position
\begin{align*}
y_l(t) = x_i(t) - x_j(t),
\end{align*}
and relative velocity
\begin{align*}
z_l(t) = v_i(t) - v_j(t), ~ t \in [t_k^q,t_{k}^{q+1})
\end{align*}
where edge $l$ is connected between agents $i$ and $j$.
Letting $x=[x_1, \cdots,x_N]^{\t}$, $v=[v_1, \cdots,v_N]^{\t} $, $ y= [y_1, \cdots,y_{M_{\sigma(t)}}]^{\t}$ and $z = [z_1, \cdots,z_{M_{\sigma(t)}}]^{\t} $, we have $y =B_{\sigma(t)}^{\t}x$ and $z =B_{\sigma(t)}^{\t}v$.

In the following, the performance specifications are imposed on tracking errors $y_l(t)$ and $z_l(t)$.
For convenience, the symbol ${\rm s}$ of section \ref{Prescribed Performance Control} is replaced with ${\rm y}$ and ${\rm q}$ respectively.

\begin{assumption} [\cite{Macellari2017Multi}] \label{assumption-3}
  The values $\hat{y}_l(t_k^q)$ and  $\hat{z}_l(t_k^q)$ are inside the performance bounds \eqref{prescribed performance}, $\forall l \in \mathcal{H}_{\sigma(t)}$.
\end{assumption}

\textit{{\fontsize{16}{0}Objective}}:
For the MAS \eqref{mas} with switching topologies $\mathcal{G}_{\sigma(t)}$ and performance functions $\rho_{y,l}(t)$ and $\rho_{z,l}(t)$,
 design a distributed control protocol $u_i(t)$ to achieve consensus in the sense of
 \begin{align*}
 \lim_{t \to \infty} y_l(t) = 0 ~{\rm and}~ \lim_{t \to \infty} z_l(t) = 0, ~ \forall l \in \mathcal{H}_{\sigma(t)}
 \end{align*}
while transient performance $\hat{y}(t) \in D_{\hat{y}}$ and $\hat{z}(t) \in D_{\hat{z}}$, $t\ge 0$ are guaranteed.



\section{ Main Results }

To achieve the above objective, we propose the prescribed performance control protocol as follows
\begin{align}\label{u1}
u_i(t)=& -\sum_{l=1}^{{M_{\sigma(t)}}}b_{il}(t)J_{y,l}(\hat{y}_{l},t)\varepsilon_{y,l}(\hat{y}_{l}) \nonumber\\
     & -\phi\sum_{l=1}^{{M_{\sigma(t)}}}b_{il}(t)J_{z,l}(\hat{z}_{l},t)\varepsilon_{z,l}(\hat{z}_{l}),\; i \in \mathcal{V}
\end{align}
where $\phi$ is a positive constant.
Invoking \eqref{mas} and \eqref{u1}, the closed-loop system is rewritten as
\begin{align}\label{compact form-1}
   \ddot{x}  = -B_{\sigma(t)}J_y(\hat{y},t)\varepsilon_y(\hat{y})- \phi B_{\sigma(t)} J_z(\hat{z},t)\varepsilon_z(\hat{z}) .
\end{align}

\begin{remark}
Compared to existing literatures \cite{Karayiannidis2012Multi, chen2019consensus, restrepo2022asymptotic} based on single-integrator MASs, the velocity information of double-integrator MAS \eqref{mas} makes controller design and stability analysis more complicated.
The control protocol \eqref{u1} consists of two nonlinear error transformation terms, which is a vital advantage in improving the transient performance of relative positions and relative velocities, instead of single relative positions mentioned in \cite{Macellari2017Multi}.
\end{remark}

Partly inspired by \cite{Hou2023Event}, the following lemmas are obtained.

\begin{lemma}[\cite{Hou2023Event, Hou2023Prescribed}]
Given a matrix
$
  Q =
  \scriptsize{
  \setlength{\arraycolsep}{0.8pt}
  \begin{bmatrix}
   h_1 I_N & -h_2 I_N \\
   h_3 I_N & h_4 I_N
  \end{bmatrix}
  }
$
with positive constants $ h_1$, $h_2$, $h_3$, $h_4 $ and any vectors $\epsilon_1, \epsilon_2 \in \mathbb{R}^N$.
If $ 4 h_1h_4 > (h_3-h_2)^2 $, then
 \begin{align}  \label{ineq-1}
   \begin{bmatrix}
   \epsilon_1  \\
   \epsilon_2
  \end{bmatrix}^{\t}
 Q
  \begin{bmatrix}
   \epsilon_1  \\
   \epsilon_2
  \end{bmatrix}
  > 0.
 \end{align}

\end{lemma}

\begin{lemma} [\cite{Hou2023Event, Hou2023Prescribed}] \label{lemma-5}
 Under Assumption \ref{assumption-3}, consider the MAS \eqref{mas} and protocol \eqref{u1}.
 If initial values $x(0)$ and $v(0)$ are bounded,
 we obtain that $x(t)$ and $v(t)$ are uniformly continuous and bounded for $t \in [0,t_{max})$.
\end{lemma}

\begin{IEEEproof}
Simplifying the equality \eqref{u1}, we obtain
\begin{align}\label{S-i-t}
  \dot{v}_i(t) = S_i(t), \; i \in \mathcal{V}
\end{align}
where
$ S_i(t) = -\sum_{l=1}^{M_{\sigma(t)}}b_{il}(t)J_{y,l}(\hat{y}_{l},t)\varepsilon_{y,l}(\hat{y}_{l})  -\phi\sum_{l=1}^{M_{\sigma(t)}}b_{il}(t)J_{z,l}(\hat{z}_{l},t)\varepsilon_{z,l}(\hat{z}_{l})$, for $ t \in [t_k^q,t_{k}^{q+1})$.
By solving differential equation \eqref{S-i-t}, one has, with $t \in [t_k^q,t_{k}^{q+1})$,
\begin{align*}
 v_i(t) =  v_i(t_k^q) + \int_{t_k^q}^{t} S_i(\tau) d\tau .
\end{align*}
Since $\hat{y}(0)\in D_{\hat{y}}$ and $\hat{z}(0)\in D_{\hat{z}}$, and $x(0)$ and $v(0)$ are bounded,
 $v_i(t)$ and $x_i(t) = x_i(0) + \int_{0}^{t_0^1} v_i(\tau) d\tau$, $t\in[0, t_0^1)$ are continuous and bounded.
By calculating, $S_i(t_0^1)$ is bounded, and thus $v_i(t)$ and $x_i(t)$, $t\in[t_0^1, t_0^2)$ are continuous and bounded.
In return, the same conclusion can be reached for $x(t)$ and $v(t)$, $t\in[t_k^q, t_k^{q+1})$.
Letting $t \to t_{max}$,
it follows from \cite{stamouli2022robust} that $x_i(t)$ and $v_i(t)$, $ t \in [0,t_{max})$ are uniformly continuous and bounded.
The proof is completed.
\end{IEEEproof}

\begin{theorem} \label{theorem-1}
Under Assumptions \ref{assumption-2} and \ref{assumption-3},
consider the MAS \eqref{mas} and protocol \eqref{u1} with performance
functions $\rho_{y,l}(t)$ and $\rho_{z,l}(t)$.
If positive constants $h_1,h_2,h_3,h_4, h_5, h_6, \phi$ are selected such that
\begin{align*}
  4 h_1h_4 > (h_3-h_2)^2, \;  h_3 - h_2 - 2h_5 \bar{\alpha}_{y} > 0,\\
  h_4\phi - h_6 \bar{\alpha}_z > 0, \; 2h_6 \phi - a_2\phi(h_3- h_2) - 2h_6 a_4 > 0,
\end{align*}
then tracking errors $y_l(t)$ and $z_l(t)$ evolve within performance bound \eqref{prescribed performance} respectively, and converge to zero as $t \to \infty$.
\end{theorem}

\begin{IEEEproof}
The proof is divided into four steps.
To begin with, there exists a unique maximum solution $w(t)= [\hat{y}(t), \hat{z}(t)]^{\t}$ over the set $D = D_{\hat{y}} \times D_{\hat{z}}$, i.e., $w(t) \in D$, $\forall t \in [0, \tau_{\max}) $.
Nextly, it is verified that the protocol $\eqref{u1}$ ensures,
$\hat{y}(t)$ and $\hat{z}(t)$ for $t \in [0, \tau_{\max})$ are bounded and strictly within the compact subset of $D_{\hat{y}}$ and $D_{\hat{z}}$ respectively.
Then proof by contradiction leads to $\tau_{\max} = \infty$.
The consensus is finally proven by Barbalat’s Lemma.

Phase \uppercase\expandafter{\romannumeral 1}.
Since $\hat{y}(0)\in D_{\hat{y}}$ and  $\hat{z}(0)\in D_{\hat{z}}$, one has $w(0) \in D$.
By calculating the derivative of $\hat{y}(t)$ and $\hat{z}(t)$, it is confirmed that $\dot{w}(t)$ is continuous and locally Lipschitz on $w$.
In view of Theorem 54 of \cite{Sontag1998Mathematical},
there exists the maximal solution $w(t)$, and hence $w(t)\in D$ is guaranteed for $t \in [0,\tau_{\max})$.

Phase \uppercase\expandafter{\romannumeral 2}.
It follows from Phase \uppercase\expandafter{\romannumeral 1} that $y_l(t)$ and $z_l(t)$, $\forall t \in [0, \tau_{\max})$ satisfy bound \eqref{prescribed performance} respectively.
Consider a potential function as follows
\begin{align} \label{potential function}
  V(t) = \frac{1}{2}\xi^{\t}Q\xi + \frac{h_5}{2} \varepsilon_y^{\t}(\hat{y}) \varepsilon_y(\hat{y})
        + \frac{h_6}{2} \varepsilon_z^{\t}(\hat{z}) \varepsilon_{z} (\hat{z} )
\end{align}
where $\xi= [ x^{\t} , v^{\t} ]^{\t} $,
$
Q =
\scriptsize{
\setlength{\arraycolsep}{0.8pt}
\begin{bmatrix}
 h_1 I_N & -h_2 I_N \\
 h_3 I_N & h_4 I_N
\end{bmatrix} > 0
}
$,
$h_5 > h_4$, and $h_6 > 0$.
Computing $\dot{V}(t)$, we obtain
\begin{align*} 
\dot{V}(t)
= & -\frac{1}{2} (h_3- h_2)y^{\t}J_y(\hat{y},t)\varepsilon_y(\hat{y}) \\
 & + h_5  \varepsilon^{\t}(\hat{y}) J_y(\hat{y},t) \alpha_{y}(t)y - h_6 \phi \|B_{\sigma(t)} J_z(\hat{z},t)\varepsilon_z(\hat{z})\|^2 \\
 & - h_4 \phi z^{\t}J_z(\hat{z},t)\varepsilon_z(\hat{z}) + h_6\varepsilon_{z}^{\t}(\hat{z})J_{z}(\hat{z},t)\alpha_{z}(t)z \\
 & + h_1 v^{\t} x - \frac{\phi}{2}(h_3- h_2)y^{\t} J_z(\hat{z},t)\varepsilon_z(\hat{z}) \\
 & + (h_5-h_4) z^{\t}J_y(\hat{y},t)\varepsilon_y(\hat{y})  +  \frac{1}{2} (h_3-h_2)  v^{\t} v  \\
 & - h_6\varepsilon_{z}^{\t}(\hat{z})J_{z}(\hat{z},t) B_{\sigma(t)}^{\t}B_{\sigma(t)}J_y(\hat{y},t)\varepsilon_y(\hat{y}) .
\end{align*}
By the Young’s inequality \cite{chen2019Second}, with positive constants $a_1, a_2, a_3, a_4$ and $a_5$,  one has
\begin{align*}
 & v^{\t}x \leq a_1 x^{\t}x + \frac{1}{4a_1} v^{\t}v, \\
 & y^{\t} J_z(\hat{z},t)\varepsilon_z(\hat{z}) \leq a_2 \|B_{\sigma(t)} J_z(\hat{z},t)\varepsilon_z(\hat{z})\|^2 + \frac{1}{4a_2} x^{\t}x, \\
 & z^{\t}J_y(\hat{y},t)\varepsilon_y(\hat{y}) \leq a_3 \|B_{\sigma(t)} J_y(\hat{y},t)\varepsilon_y(\hat{y})\|^2 + \frac{1}{4a_3} v^{\t}v,  \\
 & \varepsilon_z^{\t}(\hat{z}) J_{z}(\hat{z},t) B_{\sigma(t)}^{\t}B_{\sigma(t)}J_y(\hat{y},t)\varepsilon_y(\hat{y}) \\
 & \quad \leq a_4 \|B_{\sigma(t)} J_{z}(\hat{z},t) \varepsilon_z^{\t}(\hat{z})\|^2 + \frac{1}{4a_4} \|B_{\sigma(t)} J_y(\hat{y},t) \varepsilon_y(\hat{y})\|^2.
\end{align*}
Combining the above inequalities and inequality \eqref{baralpha} gives
\begin{align*}
\dot{V}(t)
\leq & -(\frac{h_3- h_2}{2}  -  h_5 \bar{\alpha}_{y} )y^{\t}J_y(\hat{y},t)\varepsilon_y(\hat{y})  \\
 & -(h_6\phi -\frac{a_2\phi (h_3- h_2)}{2} - h_6a_4) \|B_{\sigma(t)}J_z(\hat{z},t)\varepsilon_z(\hat{z})\|^2    \\
 & -(h_4\phi - h_6 \bar{\alpha}_z) z^{\t}J_z(\hat{z},t)\varepsilon_z(\hat{z}) \\
 & + ( h_1 a_1 + \frac{\phi (h_3- h_2) }{8a_2} ) x^{\t}x \\
 & + ( \frac{h_1 }{4a_1} + \frac{(h_5-h_4)}{4a_3} + \frac{(h_3-h_2)}{2} ) v^{\t} v  \\
 & + ( (h_5-h_4)a_3   +  \frac{h_6}{4a_4} ) \|B_{\sigma(t)} J_y(\hat{y},t) \varepsilon_y(\hat{y})\|^2 .
\end{align*}
 By feat of Lemma \ref{lemma-5}, $ x^{\t}x $ and $v^{\t}v $ are bounded and the upper bound are supposed as $D_1$ and $D_2$ respectively.
 On account of the boundedness of $y(t)$ for $t \in [0,\tau_{\max})$,
 $\|B_{\sigma(t)} J_y(\hat{y},t) \varepsilon_y(\hat{y})\|^2$ is bounded and its upper bound is recorded as $D_3$.
 With $ h_6 \phi - \frac{a_2\phi}{2}(h_3- h_2) - h_6 a_4 > 0 $, one has
\begin{align*}
 \dot{V}(t) \leq & -\kappa_1 y^{\t}J_y(\hat{y},t)\varepsilon_y(\hat{y})  + \kappa_3 D_1  + \kappa_4 D_2  \\
 &- \kappa_2 z^{\t}J_z(\hat{z},t)\varepsilon_z(\hat{z}) + \kappa_5 D_3
\end{align*}
where
$ \kappa_1 =  \frac{1}{2} (h_3- h_2) -  h_5 \bar{\alpha}_{y} $,
$ \kappa_2 = h_4\phi - h_6 \bar{\alpha}_z $,
$ \kappa_3 = h_1 a_1 + \frac{\phi (h_3- h_2) }{8a_2} $,
$ \kappa_4 = \frac{h_1 }{4a_1} + \frac{(h_5-h_4)}{4a_3} + \frac{(h_3-h_2)}{2} $, and
$ \kappa_5 = (h_5-h_4)a_3   +  \frac{h_6}{4a_4} $.
By Lemma \ref{lemma-1}, the following holds
 \begin{align*}
   - \frac{\kappa_1}{2} y^{\t}J_y(\hat{y},t)\varepsilon_y(\hat{y})   + \kappa_3 D_1 +  \kappa_4 D_2 & \leq 0, \\
   - \frac{\kappa_2}{2} z^{\t}J_z(\hat{z},t)\varepsilon_z(\hat{z}) + \kappa_5 D_3 & \leq 0,
\end{align*}
for all $\|\varepsilon_y(\hat{y})\| > \bar{\varepsilon}_y$ and $\|\varepsilon_z(\hat{z})\| > \bar{\varepsilon}_z$.
By applying the inequality \eqref{ineq vare}, we have
\begin{align} \label{potential function-final}
 \dot{V}(t)
 \leq & - \frac{\kappa_1}{2} y^{\t}J_y(\hat{y},t)\varepsilon_y(\hat{y})
     - \frac{\kappa_2}{2} z^{\t}J_z(\hat{z},t)\varepsilon_z(\hat{z}) \nonumber \\
 \leq &  - \frac{\kappa_1\zeta_{y,1} }{2} \| \varepsilon_y(\hat{y})\|^2
     - \frac{\kappa_2 \zeta_{z,1} }{2} \| \varepsilon_z(\hat{z}) \|^2 \nonumber \\
 \leq & 0.
\end{align}
It can be seen that $ \dot{V}(t) \leq 0 $ is guaranteed by
\begin{align*}
 \vert \varepsilon_{y,l}(\hat{y})\vert &\le \varepsilon_y^{*}=\max\{\varepsilon_y(\hat{y}(0)),\bar{\varepsilon}_y \}, \\
 \vert \varepsilon_{z,l}(\hat{y})\vert &\le \varepsilon_z^{*} = \max\{\varepsilon_z(\hat{z}(0)),\bar{\varepsilon}_z \}, \forall l \in \mathcal{H}_{\sigma(t)}
\end{align*}
within $t\in[0,t_{max})$.
By using the inverse operation of the transformation function $ \varepsilon_{s,l}(\hat{s}_{l}) $, we obtain, with $t\in[0,t_{max})$,
\begin{align*}
 \hat{y}_l(t) & \in [\underline{\delta}_{y,l}, \bar{\delta}_{y,l}] \triangleq [-\varepsilon^{-1}_{y,l}(\varepsilon_y^{*}), \varepsilon^{-1}_{y,l}(\varepsilon_y^{*}) ], \\
 \hat{z}_l(t) & \in [\underline{\delta}_{z,l}, \bar{\delta}_{z,l}] \triangleq [-\varepsilon^{-1}_{z,l}(\varepsilon_z^{*}), \varepsilon^{-1}_{z,l}(\varepsilon_z^{*}) ].
\end{align*}

Phase \uppercase\expandafter{\romannumeral 3}.
Up to this point, we prove that $\tau_{\max}$ can be extended to $\infty$.
It is obtained from \eqref{potential function-final} that
\begin{align*}
w(t)\in D^{'} = D_{\hat{y}}^{'} \times D_{\hat{z}}^{'}, ~ \forall t \in [0, \tau_{\max})
\end{align*}
where $D_{\hat{y}}^{'} = [ \underline{\delta}_{y_1}, \bar{\delta}_{y_1} ] \times \cdots \times [ \underline{\delta}_{y_M}, \bar{\delta}_{y_M} ] $ and $D_{\hat{z}}^{'} = [ \underline{\delta}_{z_1}, \bar{\delta}_{z_1} ] \times \cdots \times [ \underline{\delta}_{z_M}, \bar{\delta}_{z_M} ] $.
$D^{'}$ is obviously a nonempty subset of $D$.
Suppose $\tau_{\max} < \infty$, it is inferred from Proposition C.3.6 of \cite{Sontag1998Mathematical} that the inequality $w(t^{'})\notin D^{'}$ holds for a instant $t^{'}\in [0, \tau_{\max})$,
which is a clear contradiction.
From Theorem $1$ of \cite{stamouli2022robust}, one obtain $\tau_{\max} = \infty$ and $w(t)\in D^{'} \subset D$, thus $V(t)$ is finite and bounded, $\forall t > 0$.

Phase \uppercase\expandafter{\romannumeral 4}.
We finally prove that $y(t)$ and $z(t)$ converge to zero as $t \to \infty$.
Taking a part of calculation in \eqref{potential function-final} gives
\begin{align}
\dot{V}(t) \leq  - \underline{\kappa} \| \varepsilon_y(\hat{y})\|^2
 - \underline{\kappa} \| \varepsilon_z(\hat{z}) \|^2
\end{align}
where $ \underline{\kappa} =  {\min}\{\frac{\kappa_1\zeta_1 }{2}, \frac{\kappa_2 \zeta_3 }{2} \}  $.
Since  $\dot{V}(t) \leq 0$ and $V(t) > 0$, $V(t)$ is bounded and $ \lim_{t \to \infty} V(t)$ exists.
Given an infinite sequences $V(t_k)$, $k \in \mathbb{N}$, by utilizing Cauchy's Convergence Criterion,
one has, for $ \forall \epsilon > 0$, $\exists M_\epsilon \in \mathcal{Z}_+$, such that $\forall k \geq M_\epsilon$,
\begin{align}
| V(t_{k+1}) -  V(t_{k}) | <  \epsilon ~ {\rm or} ~ \Big| \int_{t_k}^{t_{k+1}} \dot{V}(s) ds \Big| < \epsilon.
\end{align}
It follows that
\begin{align}
 \int_{t_k^0}^{t_k^1} -\dot{V}(s) ds + \cdots +  \int_{t_k^{m_k-1}}^{t_k^{m_k}} -\dot{V}(s) ds  < \epsilon.
\end{align}
For each subinterval $[{t_k^q}, {t_k^{q+1}})$, $q = 0, 1, \cdots, m_k-1 $, one has
\begin{align*}
   \int_{t_k^q}^{t_k^{q+1}} -\dot{V}(s) ds
& \geq  \underline{\kappa} \int_{t_k^q}^{t_k^{jq+1}} \| \varepsilon_y(\hat{y}(s))\|^2
   + \| \varepsilon_z(\hat{z}(s)) \|^2 ds \\
& \geq  \underline{\kappa} \int_{t_k^q}^{t_k^{q}+\tau} \| \varepsilon_y(\hat{y}(s))\|^2
   + \| \varepsilon_z(\hat{z}(s)) \|^2 ds .
\end{align*}
Combining the above two inequalities gives
\begin{align*}
\epsilon > & \underline{\kappa} \sum_{q=0}^{m_k-1} \int_{t_k^q}^{t_k^{q}+\tau} \| \varepsilon_y(\hat{y}(s))\|^2
           + \| \varepsilon_z(\hat{z}(s)) \|^2 ds .
\end{align*}

Since finite switches take place during $[t_k, t_{k+1})$, the number $m_k$ is finite for $k \in \mathbb{N}$.
It implies that, with a bounded constant $ M_k > 0$,
\begin{align*}
   \lim_{t \to \infty} \int_{t }^{t +\tau} M_k (\| \varepsilon_y(\hat{y}(s))\|^2 +  \| \varepsilon_z(\hat{z}(s)) \|^2)  ds  = 0 .
\end{align*}
Since $\varepsilon_y(\hat{y}(t))$ and $\varepsilon_z(\hat{z}(t))$ are bounded, $\dot{V}(t)$ and $\ddot{V}(t)$ are bounded, $ \forall t > 0$.
Utilizing the Barbalat's Lemma \cite{Hou2023Event} gives $ \lim_{t \to \infty} \varepsilon_y(\hat{y}(t)) = 0$ and $ \lim_{t \to \infty} \varepsilon_z(\hat{z}(t)) = 0$,
and consequently, $ \lim_{t \to \infty} y(t) = 0$ and $ \lim_{t \to \infty} z(t) = 0$.
The proof is completed.
\end{IEEEproof}

\begin{remark}
Generally speaking, the controller $u_i(t)$ is updated based on the connected agents' state feedback.
As the graph switches, the neighbor set of agent will change.
By applying the control protocol \eqref{u1}, the consensus control problem caused by switching topologies is solved.
If the graph $\mathcal{G}_{\sigma(t)}$ is a tree, the edge Laplacian matrix
$ B_{\sigma(t)}^{\t}B_{\sigma(t)} $ is positive definite.
However, this positive definiteness condition is not required in the proof, such that protocol \eqref{u1} is applicable to generally connected graphs containing cycles.
\end{remark}

\begin{remark}
From Theorem~\ref{theorem-1}, it follows that asymptotic consensus is achieved with transient performance, which is more accurate than practical consensus in \cite{Karayiannidis2016AModel}.
Besides, asymptotic convergence can restrain the external disturbances and bound fluctuations caused by performance bounds \cite{chen2019Second, restrepo2022asymptotic}.
\end{remark}

\begin{figure}[h]
\begin{center}
\includegraphics[width=0.38\textwidth]{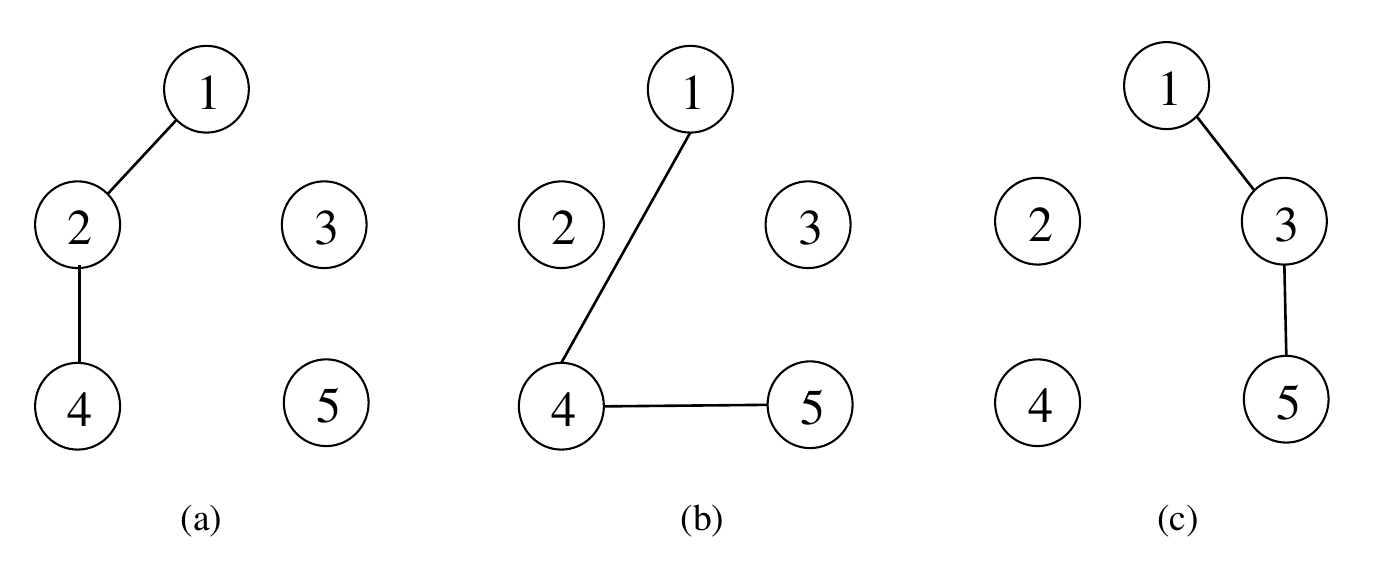}    
\caption{Topology graphs. (a) $\bar{\mathcal{G}}_1$. (b) $\bar{\mathcal{G}}_2$. (c) $\bar{\mathcal{G}}_3$. }  
\label{topology graphs}                                 
\end{center}                                 
\end{figure}

\section{Simulation}

Consider a MAS composed of $N = 5$ agents with the dynamic model \eqref{mas} and \eqref{u1}.
Suppose the topology graph described in Fig.\ref{topology graphs} switches as $ \bar{\mathcal{G}}_1 \to \bar{\mathcal{G}}_2 \to \bar{\mathcal{G}}_3 \to \bar{\mathcal{G}}_1 \to \cdots $, with the dwelling time $\tau = 0.1$ s.
We choose performance functions
$\rho_{y,l}(t) = ( 5 - 0.1 ) e^{- 1.5 t } + 0.1$ and $\rho_{z,l}(t) = ( 5 - 0.1 ) e^{- 0.8 t } + 0.1$ for all edges. Then we have
$\alpha_{y,l}(t) =  - \frac{ \dot{\rho}_{y,l}(t) }{ \rho_{y,l}(t)}= 1.5 \times \frac{ \rho_{y,l}(t)- 0.1 }{ \rho_{y,l}(t)} \leq 1.5$ and $\alpha_{z,l}(t) \leq 0.8$,
and hence $\bar{\alpha}_y = 1.5$ and $\bar{\alpha}_z = 0.8$.
For all edges, the same transformation functions are selected as
\begin{align*}
  \varepsilon_{y,l}(\hat{y}_{l})= \ln \left( \frac{ 1 +\hat{y}_{l}}{ 1-\hat{y}_{l} } \right) ~ {\rm and} ~ \varepsilon_{z,l}(\hat{z}_{l})= \ln \left( \frac{ 1 +\hat{z}_{l}}{ 1-\hat{z}_{l} } \right).
\end{align*}

\begin{figure}[tbp]
\begin{center}
\includegraphics[width=0.46\textwidth]{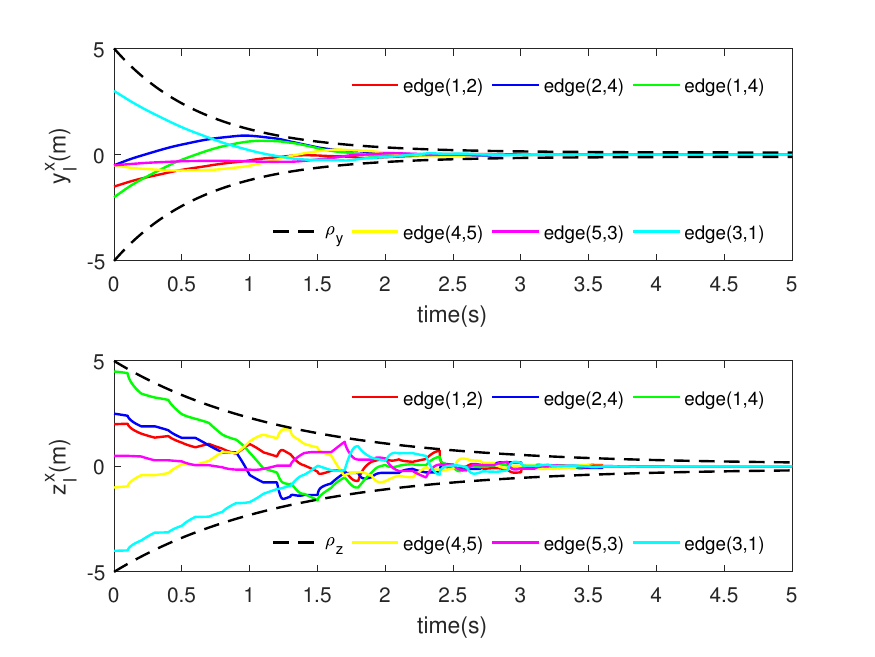}    
\caption{Convergence trajectories with prescribed performance. Top: relative positions; bottom: relative velocities. }  
\label{Convergence graphs}                                 
\end{center}                                 
%
\begin{center}
\includegraphics[width=0.46\textwidth]{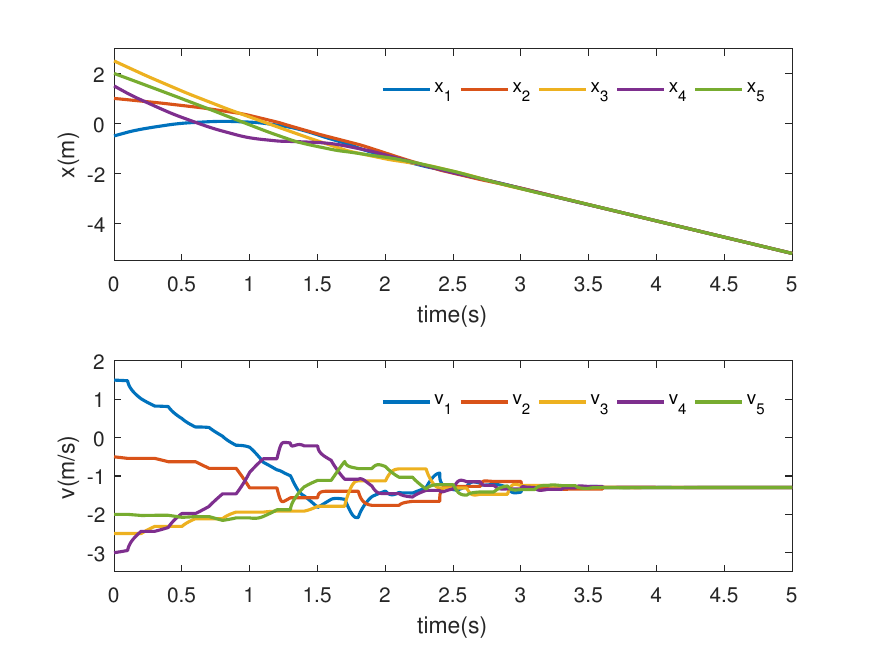}    
\caption{State trajectories of all agents. Top: position trajectories; bottom: velocity trajectories. }  
\label{Trajectory graphs}                                 
\end{center}
\end{figure}

The initial positions of agents are $x(0)=[ -0.5, \, 1, \, 2.5, \, 1.5, \, 2]^{\t}$, and initial velocities are $v(0)=[ 1.5, \, -0.5, \, -2.5, \, -3, \, -2]^{\t}$.
For Theorem~\ref{theorem-1}, we choose $h_1=10, h_2=1, h_3=6, h_4=1.5, h_5=1.6, h_6=1.5, a_2=0.1, a_3=0.5, a_4=0.1$ and $\phi = 1$.
The state trajectories of all agents for $t$ from $0$ to $5$s, are depicted in
Figs.~\ref{Convergence graphs} and \ref{Trajectory graphs}.
It follows from Fig.~\ref{Convergence graphs} that the relative positions and relative velocities are confined within the performance bounds as expected, and the state consensus of all agents is eventually achieved as shown in Fig.~\ref{Trajectory graphs}.

\section{Conclusion}

This paper addresses the full-state prescribed performance-based consensus problem for double-integrator MASs with jointly connected topologies.
A distributed prescribed performance control protocol is proposed such that the relative positions and relative velocities are constrained within the respective performance bounds.
The proposed protocol drives all agents to achieve consensus with jointly connected topologies.
Future research topics focus on more general agent dynamics.


%






\ifCLASSOPTIONcaptionsoff
  \newpage
\fi



%


\bibliographystyle{IEEEtran}
\bibliography{referencelists}

\begin{thebibliography}{10}
\providecommand{\url}[1]{#1}
\csname url@samestyle\endcsname
\providecommand{\newblock}{\relax}
\providecommand{\bibinfo}[2]{#2}
\providecommand{\BIBentrySTDinterwordspacing}{\spaceskip=0pt\relax}
\providecommand{\BIBentryALTinterwordstretchfactor}{4}
\providecommand{\BIBentryALTinterwordspacing}{\spaceskip=\fontdimen2\font plus
\BIBentryALTinterwordstretchfactor\fontdimen3\font minus
  \fontdimen4\font\relax}
\providecommand{\BIBforeignlanguage}[2]{{%
\expandafter\ifx\csname l@#1\endcsname\relax
\typeout{** WARNING: IEEEtran.bst: No hyphenation pattern has been}%
\typeout{** loaded for the language `#1'. Using the pattern for}%
\typeout{** the default language instead.}%
\else
\language=\csname l@#1\endcsname
\fi
#2}}
\providecommand{\BIBdecl}{\relax}
\BIBdecl

\bibitem{2020Networked}
X.~M. Zhang, Q.~L. Han, X.~Ge, D.~Ding, L.~Ding, D.~Yue, and C.~Peng,
  ``Networked control systems:a survey of trends and techniques,''
  \emph{IEEE/CAA Journal of Automatica Sinica}, vol. 007, no. 001, pp. 1--17,
  2020.

\bibitem{2017Recent}
J.~Qin, Q.~Ma, Y.~Shi, and L.~Wang, ``Recent advances in consensus of
  multi-agent systems: A brief survey,'' \emph{IEEE Transactions on Industrial
  Electronics}, vol.~64, no.~6, pp. 4972--4983, 2017.

\bibitem{2004Consensus}
R.~Olfati-Saber and R.~M. Murray, ``Consensus problems in networks of agents
  with switching topology and time-delays,'' \emph{IEEE Transactions on
  Automatic Control}, vol.~49, no.~9, pp. 1520--1533, 2004.

\bibitem{Yu2013Consensus}
W.~Yu, L.~Zhou, X.~Yu, J.~Lü, and R.~Lu, ``Consensus in multi-agent systems
  with second-order dynamics and sampled data,'' \emph{IEEE Transactions on
  Industrial Informatics}, vol.~9, no.~4, pp. 2137--2146, 2013.

\bibitem{zeng2022second}
J.~Zeng, P.~Wang, H.~Su, and C.~Xu, ``Second-order consensus for multi-agent
  systems with various intelligent levels via intermittent sampled-data
  control,'' \emph{IEEE Transactions on Circuits and Systems II: Express
  Briefs}, vol.~69, no.~12, pp. 4899--4903, 2022.

\bibitem{ren2021optimal}
Y.~Ren, Q.~Wang, and Z.~Duan, ``Optimal distributed leader-following consensus
  of linear multi-agent systems: A dynamic average consensus-based approach,''
  \emph{IEEE Transactions on Circuits and Systems II: Express Briefs}, vol.~69,
  no.~3, pp. 1208--1212, 2021.

\bibitem{sun2023consensus}
Y.~Sun and S.~Peng, ``Consensus of discrete-time leader-following linear
  multi-agent systems under lyapunov-function-based event-triggered
  mechanism,'' \emph{IEEE Transactions on Circuits and Systems II: Express
  Briefs}, vol.~70, no.~12, pp. 4409 -- 4413, 2023.

\bibitem{tan2020distributed}
X.~Tan, M.~Cao, and J.~Cao, ``Distributed dynamic event-based control for
  nonlinear multi-agent systems,'' \emph{IEEE Transactions on Circuits and
  Systems II: Express Briefs}, vol.~68, no.~2, pp. 687--691, 2020.

\bibitem{2008Robust}
C.~P. Bechlioulis and G.~A. Rovithakis, ``Robust adaptive control of feedback
  linearizable {MIMO} nonlinear systems with prescribed performance,''
  \emph{IEEE Transactions on Automatic Control}, vol.~53, no.~9, pp.
  2090--2099, 2008.

\bibitem{Karayiannidis2012Multi}
Y.~Karayiannidis, D.~V. Dimarogonas, and D.~Kragic, ``Multi-agent average
  consensus control with prescribed performance guarantees,'' in \emph{IEEE
  Conference on Decision \& Control}, 2012, pp. 2219--2225.

\bibitem{Macellari2017Multi}
L.~Macellari, Y.~Karayiannidis, and D.~V. Dimarogonas, ``Multi-agent second
  order average consensus with prescribed transient behavior,'' \emph{IEEE
  Transactions on Automatic Control}, vol.~62, no.~10, pp. 5282--5288, 2017.

\bibitem{Karayiannidis2016AModel}
Y.~Karayiannidis, D.~Papageorgiou, and Z.~Doulgeri, ``A model-free controller
  for guaranteed prescribed performance tracking of both robot joint positions
  and velocities,'' \emph{IEEE Robotics \& Automation Letters}, vol.~1, no.~1,
  pp. 267--273, 2016.

\bibitem{chen2019consensus}
F.~Chen and D.~V. Dimarogonas, ``Consensus control for leader-follower
  multi-agent systems under prescribed performance guarantees,'' in \emph{2019
  IEEE 58th Conference on Decision and Control}.\hskip 1em plus 0.5em minus
  0.4em\relax IEEE, 2019, pp. 4785--4790.

\bibitem{chen2019Second}
------, ``Second order consensus for leader-follower multi-agent systems with
  prescribed performance,'' in \emph{IFAC-PapersOnLine}, vol.~52, no.~20, 2019,
  pp. 103--108.

\bibitem{Hou2021Event}
Y.~Hou and W.~Hu, ``Event-triggered consensus of second-order multi-agent
  systems with prescribed performance,'' in \emph{2021 China Automation
  Congress (CAC)}, 2021, pp. 1146--1150.

\bibitem{Hou2023Event}
Y.~Hou and B.~Cheng, ``Event-based {$H_{\infty}$} consensus of
  double-integrator multi-agent systems: A prescribed performance control
  approach,'' \emph{IEEE Transactions on Circuits and Systems II: Express
  Briefs}, early access, November. 15, 2023, doi:{\color{blue}
  \href{https://ieeexplore.ieee.org/document/10318206}{10.1109/TCSII.2023.3332746}}.

\bibitem{cheng2018event}
B.~Cheng, X.~Wang, and Z.~Li, ``Event-triggered consensus of homogeneous and
  heterogeneous multiagent systems with jointly connected switching
  topologies,'' \emph{IEEE Transactions on Cybernetics}, vol.~49, no.~12, pp.
  4421--4430, 2018.

\bibitem{wan2023differentially}
X.~Wan, Y.~Guo, and X.~Wu, ``Differentially private consensus for multi-agent
  systems under switching topology,'' \emph{IEEE Transactions on Circuits and
  Systems II: Express Briefs}, vol.~70, no.~9, pp. 3499--3503, 2023.

\bibitem{stamouli2022robust}
C.~J. Stamouli, C.~P. Bechlioulis, and K.~J. Kyriakopoulos, ``Robust dynamic
  average consensus with prescribed transient and steady state performance,''
  \emph{Automatica}, vol. 144, p. 110503, 2022.

\bibitem{restrepo2022asymptotic}
E.~Restrepo and D.~V. Dimarogonas, ``On asymptotic stability of
  leader--follower multiagent systems under transient constraints,'' \emph{IEEE
  Control Systems Letters}, vol.~6, pp. 3164--3169, 2022.

\bibitem{Hou2023Prescribed}
Y.~Hou, W.~Hu, J.~Li, and T.~Huang, ``Prescribed performance control for
  double-integrator multi-agent systems: A unified event-triggered consensus
  framework,'' \emph{IEEE Transactions on Circuits and Systems I: Regular
  Papers}, 2023, under review.

\bibitem{Sontag1998Mathematical}
E.~D. Sontag, \emph{Mathematical control theory}.\hskip 1em plus 0.5em minus
  0.4em\relax London, U. K.: Springer, 1998.

\end{thebibliography}


%








\end{document}